\documentclass[epj]{svjour}
\usepackage{amsmath,amssymb}
\usepackage[dvips]{graphics,graphicx}

\begin{document}

\title{Emergence of superfluid transport in a dynamical system of
ultra-cold atoms}

\author{Joachim Brand\inst{1}%
\thanks{\emph{Present address:} Institute of Fundamental Sciences,
Massey University (Albany Campus), Private Bag 102904, North Shore
MSC, Auckland, New Zealand }%
 \and Andrey R. Kolovsky\inst{1,2}}
\institute{
  \inst{1} Max Planck Institute for the Physics of Complex
Systems, N\"othnitzer Stra{\ss}e 38, 01187 Dresden, Germany\\
  \inst{2} Kirensky Institute of Physics, 660036 Krasnoyarsk, Russia}
\date{Received: date / Revised version: date}

\abstract{
The dynamics 
of a Bose-Einstein condensate is studied theoretically in
a combined periodic plus harmonic external potential. Different
dynamical regimes of stable and unstable collective dipole and Bloch
oscillations are analysed in terms of a quantum mechanical pendulum
model. Nonlinear interactions are shown to counteract
quantum-mechanical dephasing and lead to phase-coherent, superfluid transport.
}

\maketitle

\section{Introduction}
\label{sec1}

The study of transport properties of ultra-cold atoms in corrugated
potentials has become an intensely discussed topic since the first
experiments with Bose-Einstein condensates (BECs) in optical lattices
almost a decade ago
\cite{Anderson1998a,cataliotti2001,Burger2001a,Greiner2002a,Smerzi2002a,cataliotti03,cataliotti03a,orso:020404,pezze:120401,ott:160601,fertig:120403,rey:033616,altman:020402,polkovnikov:063613}.
The observed or predicted phenomena are often discussed by concepts
borrowed from the phenomenology of extended systems. This approach led
to the characterisation of superfluid and insulating phases and phase
transitions \cite{Greiner2002a,altman:020402}, modulational
instability\cite{Wu03,Konotop2002a}, and dissipative
behaviour\cite{Burger2001a,polkovnikov:063613}. However, different dynamical
regimes like small-amplitude oscillations, dephasing instabilities,
and Bloch oscillations require different models and analogies for
their explanation.

In this work we approach the problem from a different, somewhat
{\em holistic} point of view and treat the BEC in the external potential as
a finite dynamical system. By mapping this problem onto a simple
pendulum model, we are able to explain different dynamical regimes as
well as stabilisation and destabilisation mechanisms in a unified approach.
Specifically, we consider a cloud of ultra-cold bosonic atoms
in a one-dimensional (1D) optical lattice in a classical
tight-binding approximation with additional harmonic trapping in the
lattice axis. We further confine the analysis to the situation of a
sufficiently large number of atoms per site that quantum fluctuations
may be neglected . Typical experiments with hundreds of
atoms on the central site certainly satisfy this condition but
probably the theory still remains valid with much lower atom numbers.
This system has been discussed before in many
experimental and theoretical works. A dynamical ``superfluid-insulator''
transition has been predicted in Ref.~\cite{Smerzi2002a} for a BEC on
the basis of a modulational instability caused by nonlinear
interactions. A similar effect of insulating behaviour, however, was
observed in noninteracting fermions \cite{pezze:120401}. The latter
was interpreted as a very different mechanism in terms of a
semiclassical pendulum model. Experiments with BECs
\cite{cataliotti03,cataliotti03a} and further numerical analysis
\cite{adhikari03,nesi04} gave ambiguous results in showing reduced
mobility of bosons without revealing the mechanism.

In this paper we study the BEC in a combined harmonic and lattice trap
by exploiting a mapping of this system to a quantum mechanical
pendulum model. This exact mapping of the lattice dynamics of the
noninteracting system to a simple quantum pendulum model establishes
two effects: A separatrix in the semiclassical phase space leads to
two separate regions of qualitatively different dynamics (see
Fig.~\ref{fig:PhaseSpace} and discussion in
Sec.~\ref{sec2}). Furthermore, dephasing occurs due to
quantum-mechanical wave packet motion. Adding non-linear interactions
introduces two additional effects: Far away from the separatrix, the
nonlinearity counteracts the quantum dephasing of the linear problem
and thus stabilises wave packet motion.
We understand the emerging coherent wave-packet dynamics 
as a signature of superfluid transport in the sense discussed in the
recent literature (see, e.g.\ Refs.\
\cite{Smerzi2002a,cataliotti03,ott:160601,fertig:120403}). 
Close to the separatrix and depending on the
strength of the nonlinearity, a dynamical instability destroys
coherent wave packet motion. We thus obtain a unified view of such
different phenomena as dynamical instability, dephasing, and
superfluidity.  While a dynamical instability has been predicted by a
different mechanism in Ref.~\cite{Smerzi2002a} and the effect of the
separatrix has been discussed in conjuction with fermions in
Ref.~\cite{pezze:120401}, we believe that quantum dephasing and its
suppression by nonlinear interactions has not been discussed
before\cite{note-dephasing}.  In addition to the stabilisation of
dipole oscillations, we also predict a so far undescribed regime of
coherent wave packet motion above the separatrix. This regime could be
exploited to generate the recently proposed atomic gas at negative
kinetic temperatures \cite{mosk:040403}.

\section{Non-interacting atoms}
\label{sec2}

We begin with dynamics of non-interacting atoms, governed
by the Schr\"odinger equation with the following single-particle
Hamiltonian
\begin{equation}
\label{1a}
\widehat{H}=\frac{\hat{p}^2}{2 M}-V_0\cos^2\left(\frac{2\pi}{d}x\right)
+\frac{M\omega^2}{2}x^2 \;.
\end{equation}
In Eq.~(\ref{1a}) $M$ is the atomic mass, $V_0$ the depth of the
optical potential, $d$ the lattice period, and $\omega$ the frequency of
the harmonic confinement.  If the lattice potential is large compared
to the recoil energy $E_R=2 \hbar^2 \pi^2/d^2 M$, we may use the
tight-binding ansatz for the Schr\"odinger equation, $\psi(x,t)=\sum_l
a_l(t)\psi_l(x)$, where $\psi_l(x)$ are the localised Wannier
states. This leads to a system of coupled linear equations for
the complex amplitude $a_l(t)$,
\begin{equation}
\label{1}
i\hbar\dot{a}_l=\frac{\nu}{2}l^2 a_l-\frac{J}{2}\left(a_{l+1}+a_{l-1}\right) \;,
\end{equation}
where $\nu=M\omega^2d^2$ and $J$ is the hopping matrix element,
uniquely defined by the depth of the optical lattice $s=V_0/E_R$
as $J/E_R \sim s^{3/4} \exp(-2 \sqrt{s})/\sqrt{\pi}$.
\begin{figure}[t]
\center
\includegraphics[width=8.5cm, clip]{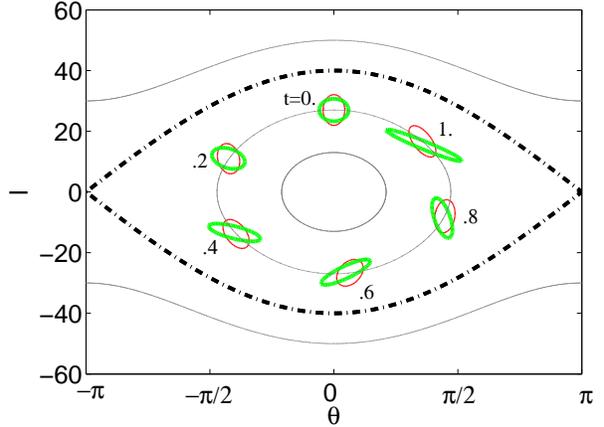}
\caption{Phase space of the classical pendulum in the variables $l$
and $\theta$. The dash-dotted lines show the separatrix for $l^*=40$,
separating rotations from oscillations. These correspond to Bloch and
dipole oscillations of a BEC, respectively. Thin (grey) lines indicate
classical trajectories with initial conditions $\theta_0=0$ and $ l_0
= 13, 27, 50, -50$.  The quantum time evolution of a Gaussian
wave packet is schematised by the 50\% contour line of the
corresponding Wigner function \cite{schleich01book}. After initial
displacement from equilibrium position to $l_0 =27$, $\theta_0=0$,
solutions of Eqs.~(\ref{7}) are shown in anti-clockwise order for the
quantum pendulum at $g=0$ (thick green contours) and the interacting
case at $g/J=1$ (thin red contours) at times as indicated.}
\label{fig:PhaseSpace}
\end{figure}

\begin{figure}[t]
\center
\includegraphics[width=\columnwidth,clip]{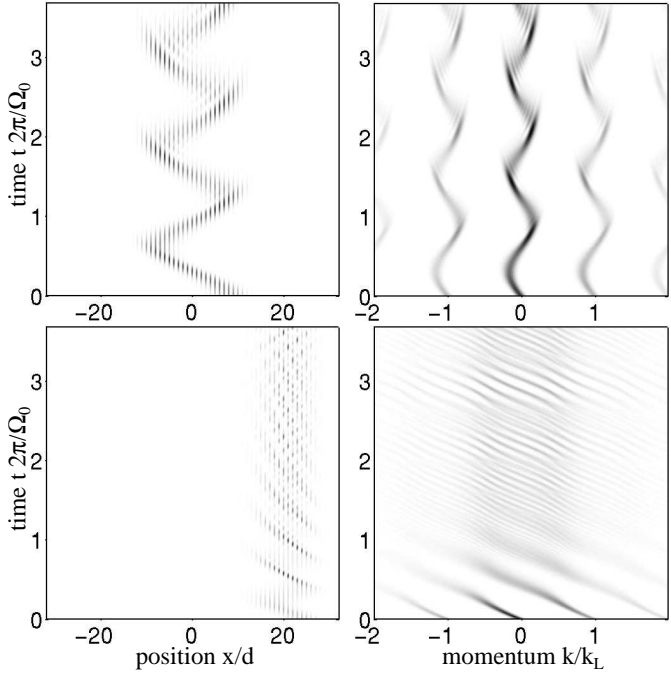}
\caption{Dynamics of non-interacting atoms
($g=0$). Gray scale images show the time evolution of the squared wave
function after the ground state of the system has been suddenly
displaced by the distance $\Delta x$ at $t=0$. The left and right
columns show real space and  momentum space
plots, respectively, for $\Delta x/d=8$ (dipole 
oscillations) in the upper row and for $\Delta x/d=24$
(Bloch oscillations) in the lower row.
The lattice parameters correspond to $J=2.4\cdot 10^{-2}E_R$,
$\nu=3.2\cdot 10^{-4}E_R$ (hence, $l^*=17$).
The time axis is scaled by the period $2\pi/\Omega_0$ of
small-amplitude pendulum oscillations. Momentum is scaled by the
reciprocal lattice constant $k_L= 2\pi/d$. The multiple peak structure of the
momentum-space plots is due to the presence of a periodic potential.}
\label{fig2}
\end{figure}

A particularly transparent description of the dynamics governed by
Eq.~(\ref{1}) is obtained by mapping it to the mathematical pendulum
\cite{kolovsky04:Review}.  Indeed, 
introducing the function
$\phi(\theta,t)=1/\sqrt{2\pi}\sum_l a_l(t)\exp(il\theta)$, the system
of Eq.~(\ref{1}) reduces to the Schr\"odinger equation for the
quantum pendulum with the Hamiltonian
\begin{equation}
\label{2}
\widehat{H}=\frac{\nu}{2}\widehat{L}^2 - J\cos(\theta) \;,\quad
\widehat{L}=-i\frac{\partial}{\partial \theta} \;.
\end{equation}
This problem is related to the Mathieu
equation \cite{mattis86,hooley:080404,rigol:043627,ruuska04},
which is solved by well-known special functions \cite{abramowitz72}. The
full advantage of the representation (\ref{2}), however, is the easily
accessible interpretation in terms of pendulum dynamics.
A characteristic feature of the classical pendulum is the existence of
a particular trajectory -- the separatrix, which separates the vibrational
and rotational regimes of the pendulum, see Fig.~\ref{fig:PhaseSpace}.
The notion of the separatrix can be well extended into the quantum problem
\cite{berman81}. It is associated with the critical angular momentum,
or the critical site index of the original problem,
\begin{equation}
\label{3}
l^*=2(J/\nu)^{1/2} \;.
\end{equation}
The existence of a critical  $l^*$ has been indicated in
laboratory experiments \cite{cataliotti03},  where
the authors excite the system by suddenly shifting
the harmonic trap by a distance $\Delta x$.
Then, for $l_0=\Delta x/d<l^*$ (for the lattice parameters used in the 
cited experiment $l^*=134$) the wave packet oscillates around the
trap origin, while for $l_0>l^*$ it sticks to one side of
the parabolic potential and the centre-of-mass position can never
reach the equilibrium position. For $g=0$ 
these dynamical regimes are illustrated in Fig.~\ref{fig2}.
The characteristic frequency of the wave-packet oscillation is
given by the pendulum frequency $\Omega(l)$ \cite{lichtenberg83book}. Below the
separatrix ($l\ll l^*$) we have $\Omega(l)\approx\Omega_0 \equiv (\nu
J)^{1/2}/\hbar= 
\omega(M/M^*)^{1/2}$, where $M^*$
is the effective mass 
of an atom in the lowest Bloch band. At the separatrix
$\Omega(l^*) = 0$ and above ($l\gg l^*$) we have  $\Omega(l)\approx
\nu l/\hbar$. 
Note that the dynamics of the atoms for $l>l^*$ can be viewed as Bloch
oscillations of the atoms in a (local) static field $F=\nu l_0/d$ with
the Bloch frequency $\Omega_{\rm BO} = dF/\hbar$.

\begin{figure}[t]
\center
\includegraphics[width=\columnwidth,clip]{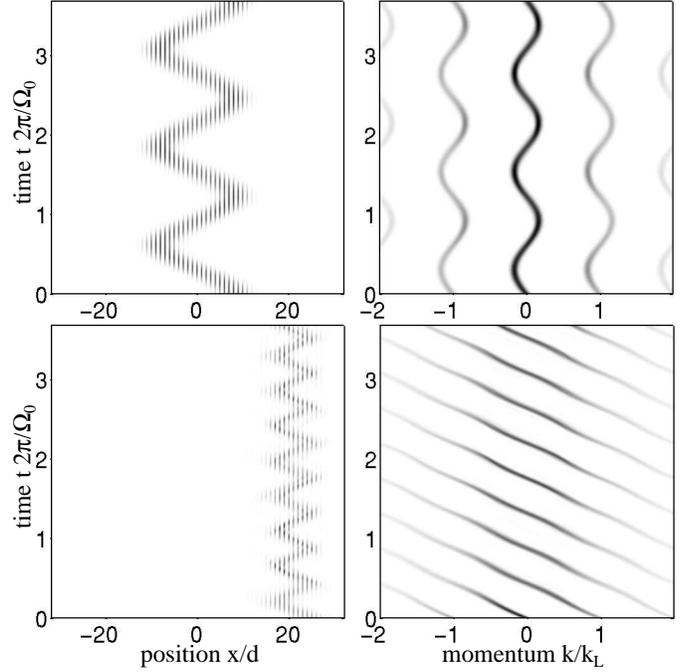}
\caption{The same as in Fig.~\ref{fig2} but for finite  nonlinearity
  $g=1.55\cdot 10^{-2}E_R$.}  
\label{fig2a}
\end{figure}

In addition to the effect of the separatrix one can also see the effect of
dephasing in Fig.~\ref{fig2}, which smears out the oscillations of the wave
packet as time goes on. This can be related to
the non-equidistant spectrum of the quantum pendulum, which is
inherited from the nonlinear frequency dependence $\Omega(l)$ of the
classical pendulum.
If $l^*\gg 1$, a short-time description of the dephasing can be
obtained by solving the equations of motion of the classical pendulum
for an {\em ensemble of trajectories} with initial conditions scattered
over the phase volume $\sim 2\pi\hbar_{\rm eff}$ with $\hbar_{\rm
  eff}=(\nu/J)^{1/2} = 2/l^*$. 
As a result we obtain a $t^2$-exponential decay for the oscillations of
the mean coordinate and momentum of the atoms.
The dephasing is conveniently quantified in terms of the quantity 
\begin{equation}
\label{3a}
\Psi =\sum_l a_l a_{l+1}^* \;,
\end{equation}
which has been introduced as an ``order
parameter'' in Ref.~\cite{Smerzi2002a}. In fact, due to the
normalisation condition $\sum_l a_l a_{l}^* = 1$, we find $\Psi\approx
1$ when the site-to-site phase fluctuations are small and $\Psi\approx
0$ in the presence of strong phase fluctuations. The upper left panel of
Fig.~\ref{figMap} shows the decay of $\Psi$ during dipole oscillations
due to dephasing.

Concluding this section we note that the calculations were
done by using the continuous nonlinear Schr\"odinger equation with
a lattice depth of $s=12.16$. For this depth of the optical
potential and the considered initial displacement, the results obtained within
the tight-binding approach (not shown) practically coincide with the depicted 
ones. The deviation between the solutions appears only for
the initial shift $l_0$ larger than $l_{max}\approx\Delta/\nu$,
where $\Delta$ is the energy gap between the ground and first excited
Bloch bands. If $l_0>l_{max}$ the Landau-Zener tunnelling takes
place and the single-particle dynamics of the atoms is a superposition
of the Bloch and dipole oscillations.

\section{Interacting atoms}
\label{sec3}

We shall analyse the case of interacting atoms in the frame of
the 1D Gross-Pitaevskii equation,
\begin{equation}
\label{1b}
i\hbar\frac{\partial \psi(x,t)}{\partial t}=\widehat{H}\psi(x,t)
+g_{1D}|\psi(x,t)|^2\psi(x,t) \;, 
\end{equation}
where $g_{1D}\sim a_s\hbar\omega_\perp N$, 
$a_s$ is the $s$-wave scattering length, $\omega_\perp$ the radial
frequency and $N$ the total number of atoms.
The tight-binding version of (\ref{1b}) reads as
\begin{equation}
\label{2a}
i\hbar\dot{a}_l=\frac{\nu}{2}l^2 a_l-\frac{J}{2}\left(a_{l+1}+a_{l-1}\right)
+g|a_l|^2a_l \;,
\end{equation}
where $g=g_{1D} \int |\psi_l(x)|^4 d x \sim a_s\hbar\omega_\perp N/d$.

\begin{figure}[t]
\center
\includegraphics[width=8.5cm, clip]{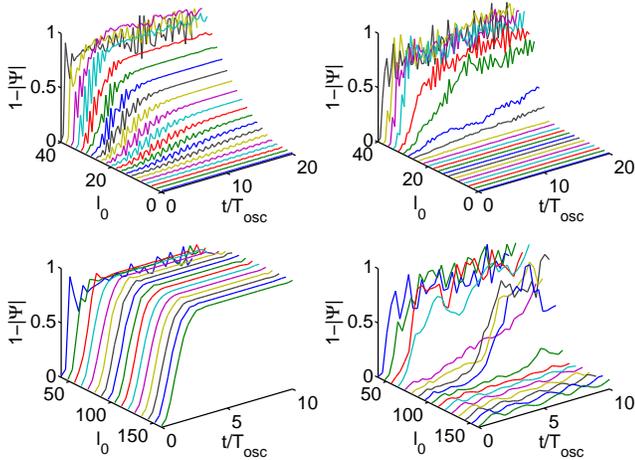}
\caption{Decay of the order parameter $\Psi$ over time for different
initial displacements $l_0$ below $l^*=40$ (upper panels) and above
$l^*$ (lower
panels). The panels on the left hand side correspond to the
noninteracting case ($g=0$) and the panels on the right to $g/J=1$.}
\label{figMap}
\end{figure}

The main result we want to report in this work is that a weak
nonlinearity can compensate the dephasing and the wave packet follows
the classical trajectory of the pendulum without dispersion.
This is illustrated in Fig.~\ref{fig2a} and  the right column of
Fig.~\ref{figMap}. 

\subsection{Variational approach}
\begin{figure}[h!]
\center
\includegraphics[width=8.5cm, clip]{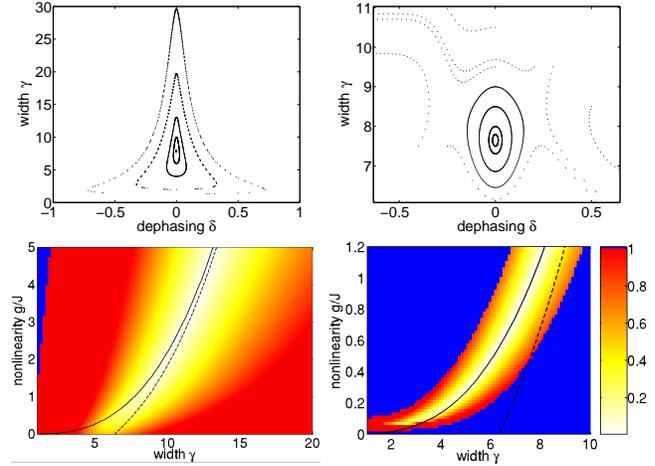}
\caption{Upper panels: Poincare cross section of the effective system
(\protect\ref{5}) for $l^*=40$ in the oscillating regime with $l_0=
l^* /2$ (left) and the rotating regime with $l_0= 4 l^*$ (right). The
periodic orbit is located in the centre of the stability island.
Lower panels: The stability region of the depicted periodic orbit in
the $(\gamma,g)$--plane. The relative deviation $ \overline{\Delta
\gamma}/\overline{\gamma}$ of the width $\gamma$ averaged over the trajectory
is shown in grey scale (according to the colorbar).  Additionally, the
solid line indicates the balance equation (\protect\ref{8}) and the
broken line shows the equilibrium width of the ground state BEC in the
given potential.}
\label{fig4}
\end{figure}

In order to estimate the amount of nonlinearity required to convert
the quantum dynamics of the pendulum into the `classical dynamics', we use
the Gaussian variational ansatz of Ref.~\cite{trombettoni01}. For the quantum
pendulum this amounts to a semiclassical approximation that can
account for dephasing \cite{heller:1544}. In this approach, the wave packet
is parametrised as
\begin{equation}
\label{4}
a_l(t)=\sqrt{A}\exp\left[-\frac{(l-L)^2}{\gamma^2}+i\theta(l-L)
+i\frac{\delta}{2}(l-L)^2\right],
\end{equation}
where $A$ is a normalisation constant.
Then the centre of the wave packet $L(t)$, the dispersion $\gamma(t)$,
the velocity $\theta(t)$, and the dephasing parameter $\delta(t)$
satisfy Hamilton's equations for the effective Hamiltonian
\begin{equation}
\label{5}
H_{\rm eff}=\frac{\nu}{2}\left(L^2+\frac{\gamma^2}{4}\right)
-J\cos\theta e^{-\eta}+\frac{g}{2\sqrt{\pi}\gamma} \;,
\end{equation}
where $\eta=1/2\gamma^2+\gamma^2\delta^2/8$ and the pairs of canonical
variables are $(L,\theta)$ and
$(\gamma^2/8,\delta)$,
respectively. Thus we have
\begin{align} 
\nonumber
\hbar \dot{L}=&-J\sin\theta e^{-\eta} \;,\quad \hbar \dot{\theta}=\nu L
\;,\quad \hbar \dot{\gamma}=J\gamma\delta\cos\theta e^{-\eta} ,\\
\label{7}\hbar \dot{\delta}=&J\cos\theta\left(\frac{4}{\gamma^4} -
\delta^2\right)e^{-\eta} +\frac{2g}{\sqrt{\pi}\gamma^3}-\nu \;.
\end{align}
The non-dispersive dynamics of the wave packet depicted
in the lower row of Fig.~\ref{fig2} implies the (quasi)periodic
dynamics of the 
variables $L$, $\theta$, $\gamma$ and $\delta$. In fact, for the
certain range of the nonlinearity $g$ and harmonic confinement $\nu$, there is
a stable periodic orbit in the four-dimensional phase space of the
system (\ref{5}), which comes through the point $\delta=0$.
The condition for the existence of this periodic orbit is approximately
given by the condition
\begin{equation}
\label{8}
g/\nu=\sqrt{\pi}\gamma^3/2\;,
\end{equation}
which means that the last two terms in the equation for $\delta$ cancel
each other.
Examples of the discussed stable periodic orbits
are shown in in the upper two panels of Fig.~\ref{fig4}
In lower panels of Fig.~\ref{fig4} we plot the stability regions of the orbit
together with the estimate (\ref{8}). The bright regions correspond
to the quasiperiodic dynamics with small deviations of $\delta$ and $\gamma$. 
In the grey (red) regions the deviations are large and in the dark
(blue) regions the dephasing $\delta$
increases without bounds. It is worth noting that the variational
ansatz (\ref{4}) becomes invalid as soon as the orbit is 
unbounded or badly bounded.
On the other hand, if $\delta(t)$ is captured around $\delta=0$
and $\gamma$ is not too small as it occurs near the centre of the
stability island, 
we have $\exp[-\eta(t)]\approx 1$ and Eq.~(\ref{7}) 
reduces to the equation of motion for a classical pendulum.

Let us estimate the minimum strength of nonlinearity needed in order
to suppress dephasing. A coarse estimate may be obtained by the
requirement that the equilibrium width of the wave packet is
compatible with Eq.\ (\ref{8}). The minimum value $g_{\rm min}$ can be
found by requiring that the width $\gamma$ obtained from Eq.~(\ref{8}) is
equal to the non-interacting equilibrium width from setting
$\dot{\delta} = 0$ and $g=0$ in Eq.~(\ref{7}). We obtain the condition
\begin{equation} \label{eqn:gmin}
  g\ge \frac{\sqrt{\pi}\nu^{7/4}}{2^{5/2} J^{3/4}}
  \approx 0.31 \left( \frac{\nu^7}{J^3}\right)^{1/4} .
\end{equation}

\subsection{Numerical results}
The above approach to the wave packet dynamics, which is based on the
effective Hamiltonian (\ref{5}), may be still oversimplified.  For
this reason and in order to check the estimate (\ref{eqn:gmin}) we run the
DNLS for different values of the nonlinearity $g$ and harmonic
confinement $\nu$. In order to reduce the number of independent
parameters we have also assumed that the shape of the initial wave packet
is defined by the ground state of the BEC before shifting the trap
centre and, hence, the wave packet width $\gamma$ is no more an
independent parameter (see the dashed line in the stability diagrams
in Fig.~\ref{fig4}).  The results of these numerical studies can be
summarised as follows:  
\begin{itemize}
\item[(i)] The effect of stabilisation is sensitive
to the initial shift of the packet $l_0$ relative to the position of
the separatrix (\ref{3}).  In particular, no stabilisation was
observed for $l_0\approx l^*$. This is actually not surprising -- the
separatrix is the most fragile trajectory of the pendulum and any tiny
perturbation destroys it. 

\item[(ii)] If $l_0$ is sufficiently far away from
the separatrix, there is a finite interval of nonlinearity $g_{\rm
min}<g<g_{\rm max}$ where the BEC oscillations are not decaying. 

\item[(iii)]
The lower boundary $g_{\rm min}$ is defined by the condition for
appearance of a (non-negligible) stability island for the effective
system (\ref{4}) and is approximately given by Eq.~(\ref{eqn:gmin}).  

\item[(iv)]
The upper boundary $g_{\rm max}$ strongly depends on $l_0$ and
sometimes is not well defined in the sense that for a large $g$ we
find a transient or incomplete stabilisation.
An example is seen in  the lower right panel of
Fig.~\ref{figMap} in the rapid increase of $1-|\psi|$ for initial
values near $l_0\approx 80$.
This result (taken
together with the existence of the stability island) suggests, that
along with the stabilisation, the nonlinearity induces a different
process in the system which destroys the regular oscillations of the
condensate when $g$ exceeds some critical value. A more sophisticated
approach than the variational ansatz (\ref{5}) is required
to take this effect into account. 
\end{itemize}

A boundary for the stability of dipole oscillations was proposed
before in Ref.~\cite{Smerzi2002a} by a simple argument based on the
modulational instability of plane wave states. This argument lead to a
critical value of $l^*/\sqrt{2}$ for the initial displacement. Our
numerical calculations loosely support this estimate for $g/J\approx
1$ (see the upper right panel on Fig.~\ref{figMap}) but also show
additional dependence on $g/J$ and $l^*$ as well as significant
deviations in other parameter regimes. It is important to realize,
however, that instable motion below the separatrix (for $l_0 <l^*$)
and the associated dephasing leads to a mean position of the wave
packet at the equilibrium position of the harmonic potential whereas
the stable or unstable motion above the separatrix (for $l_0 >l^*$) is
characterised by a nonzero offset from the equilibrium position. Hence
both effects have very different character.

\subsection{Relation to Bloch oscillations of homogeneous BEC}
\begin{figure}
\center
\includegraphics[width=8cm, clip]{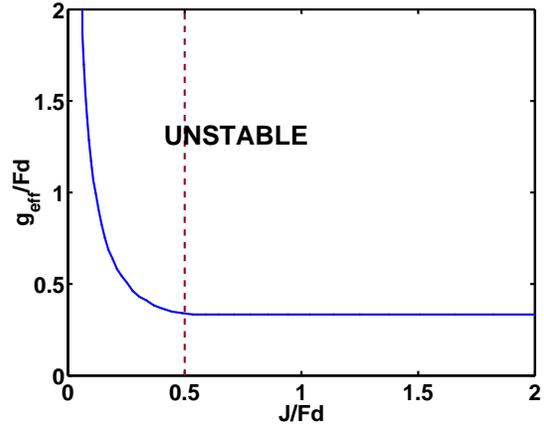}
\caption{Stability diagram for Bloch oscillations in a homogeneous
system. Here, $g_{\rm eff} = g N/L$ is an effective coupling constant.
The dashed line separates the ``universal'' regime of weak forcing,
where $g_{cr}=3.0 Fd$, 
from the ``non-universal'' regime of strong forcing, where $g_{cr}$ 
additionally depends on $J$.}
\label{fig5}
\end{figure}
At this point we briefly mention the related problem
of the dynamical (modulational) instability
that has been studied in the context of the Bloch oscillations of a
BEC subjected to a static force $F$. In this case the mean-field
equation corresponding to Eq.\ (\ref{2a}) reads
\begin{equation}
\label{eqn:statF}
i\hbar\dot{a}_l = -\frac{J}{2}\left(a_{l+1}+a_{l-1}\right) +
g|a_l|^2a_l  + F d\, l\, a_l \;, 
\end{equation}
where $d$ is the lattice constant and the initial particle number
density $|a_l|^2 = N/L$ is assumed to be constant and large compared
to one; $N$ and $L$ are the number of particles and system size,
respectively. Time-periodic solutions of Eq.\ (\ref{eqn:statF})
correspond to Bloch oscillations and generalise the Bloch oscillations
known from non-interacting particles in periodic potentials.  As shown
in the recent papers \cite{zheng:230401,preprint}, there is a critical
value of the nonlinearity, below which these Bloch oscillations are
stable, while above they are subject to a dynamical instability. This
instability scrambles the relative phases and leads to inhomogeneous,
time-aperiodic density distributions.  The stability diagram of
Fig.~\ref{fig5} summarises the results.

Since the dynamics of
the atoms in a parabolic lattice for $l_0\gg l^*$ can be alternatively
viewed as Bloch oscillations in a static field with the local
magnitude $F=\nu l_0/d$, a rough estimate for the expected instability
regime may be drawn from this critical nonlinearity.  However, an
important difference between the two systems is that the modulational
instability analysis assumes a uniform state
$\gamma\rightarrow\infty$, while in the pendulum dynamics the
finiteness of $\gamma$ is a crucial ingredient. Indeed, numerical
explorations indicate that the parameter dependence is more
complicated, which makes this an interesting problem for further
study.

\subsection{Relation to superfluidity}

The concept of superfluidity rest on a rich phenomenology rather than
precise definitions \cite{Leggett1999a}. As mentioned above, we
associate the emergence and breakdown of coherent dynamics with
superfluidity in this paper in alignment with discussions in the recent
literature related to cold atom experiments. In contrast to the
traditional approach from condensed matter theory considering infinite
systems in the thermodynamic limit, we are dealing here with an
intrinsically finite dynamical system for which the concept of
superfluidity yet has to be defined. 

Clearly, a mere application of concepts borrowed from the theory of
infinite systems will not help here: As an example we mention the
Landau critical velocity, which is bounded from above by the speed of
sound $v_s$, which is a function of the density. In a
local-density-type argument, we may consider the variation of the
density over the wave packet and conclude that the critical velocity
for the breakdown of superfluidity should vanish, as $v_s \to 0$ in
the tails of the wave packet. Thus, we would not expect superfluid
transport even though we implicitly assume that
 the atoms are completely Bose
condensed by using Eq.~(\ref{1b}).
Nevertheless, we predict coherent transport in certain
parameter regimes as discussed above. 

A systematic study of the robustness of the superfluid behaviour
against energy dissipation from small impurities is beyond the scope
of the current paper but will constitute an interesting extension of
the present work.

\section{Conclusions}
\label{sec4}

In conclusion, we have considered the 1D dynamics of a BEC of cold
atoms in parabolic optical lattices.  When interactions are absent,
this system realizes the quantum pendulum (\ref{2}) with the
experimentally controllable effective Planck constant $\hbar_{\rm
eff}=2/l^*$, where $l^*$ of Eq.~(\ref{3}) characterises the pendulum
separatrix.
The parameter $l^*$ plays an important role both in theory and
experiment. In particular, the relation between $l^*$ and the trap
centre shift $l_0=\Delta x/d$, used in the experiments to put the atoms
in motion, defines whether BEC oscillations are symmetric with respect
to the trap origin or not. The parameter $\hbar_{\rm eff}=2/l^*$
also defines the
rate of dephasing, because of which BEC oscillations decay even in the
absence of atom-atom interactions. The effect of the latter on the
discussed dynamics appears to be nontrivial. Naively, one would expect
that any nonzero interaction 
enhances the decay of BEC oscillations. However, this is not the case
-- a moderate nonlinearity is found to stabilise the oscillations,
which now can be described in terms of the {\em classical}
pendulum. The emergence of superfluid behaviour is thus related to a
quantum-classical transition. We believe that for $l_0<l^*$ the stable
regime of wave packet dynamics has been actually realized in the
experiment \cite{cataliotti03}, where periodic oscillations of a BEC
with a frequency given by the frequency of the classical pendulum have
been observed and interpreted as a superfluid phenomenon. In order to
see the transition to dephasing-dominated dynamics, the experiments
would have to work at lower particle number densities or reduce the nonlinear
coupling constant, e.g.~with $^7$Li atoms \cite{Khaykovich2002a}, by
tuning the atomic scattering 
length by means of a magnetic Feshbach resonance.  Most surprisingly,
stabilisation of wave packet motion may also occur above the
separatrix, which appears to not have been observed yet.\\[-5mm]


 \end{document}